\documentclass[a4paper,12pt]{article}
\pdfoutput=1 
\usepackage{datetime}
\usepackage{jcappub}

\usepackage[T1]{fontenc} 
\usepackage{caption}
\usepackage{subcaption}

\title{  Gauge Vectors-Tensor Gravity}

\author{Qasem Exirifard}

\affiliation{
Physics department, K. N. Toosi University of Technology,  
Tehran, Iran}
\affiliation{
Physics school, Institute for Research in Fundamental Sciences (IPM), 
Tehran, Iran
}

\emailAdd{qxfard@kntu.ac.ir}
\emailAdd{exir@theory.ipm.ac.ir}

\abstract{We review and extend the Gauge Vectors-Tensor gravity: a covariant theory of gravity composed of a metric and gauge fields, leading to simple second order partial differential equations of motion, whose Newtonian and strong limits coincide to those of the Einsten-Hilbert action but the physics of its very weak fields should be identified through observation. 

We show that GVT  is at least as dynamically  stable as the Einstein-Hilbert gravity. It accommodates the MOND paradigm.  We  study its gravitational light deflection. We show that the post Newtonian parameter of $\gamma$ vanishes in the MOND regime of GVT gravity. Since $\Lambda$CDM assumes that $\gamma=1$, this suggests to observationally measure  the $\gamma$ parameter in the weak regime of gravity as either a test for $\Lambda$CDM or  GVT models.}

\begin{document}

\maketitle
\flushbottom
\section{Introduction}
We do not know what  the essence  of our space-time is but we assume that the degrees of freedom of a geometry govern its appropriately-large-scale dynamics.  The current standard models additionally presume that the physical space-time geometry coincides to  the Riemann geometry wherein  a metric  governs the dynamics. The standard models suffer from the problem of quantum gravity. The  distribution of the known matter also can not reproduce the very large scale dynamics of the space-time in the standard models: the problem of dark matter and energy.  These problems may signal that the space-time is still a Riemannian geometry but  some other fields in addition to the metric contribute to  gravity. In other words the physical metric might be a composite field. The TeVeS model \cite{Bekenstein:2004ne} and its generalizations \cite{Sanders:2005vd} follow this possibility  and introduce a metric,  some  vectors and  scalars   in order to account for some of the observations assigned to the dark matter. We investigate  another possibility: that the physical space-time geometry may not be Riemannian.  In so doing we assume that the orbit of a massive particle is derived from the variation of 
\begin{equation}\label{s0}
S[x, \dot{x}] = m \int^p_q d\tau \left(\frac{1}{2} g_{\mu\nu} \dot{x}^\mu \dot{x}^\nu\,-\bar{A}_\mu  \dot{x}^\mu \right)\,,
\end{equation}
with respect to $x^\mu$ where $\bar{A}_\mu$ is a gauge field, and $\tau$ is an affine parameter. In other words we consider that the physical geometry can be Finslerian of the Randers type  \cite{Randers,Exirifard:2011bz}.

\section{Metric's and matters' actions}
We should recover the Einstein-Hilbert gravity in the strong and Newtonian regimes of the theory. So we demand that  
\begin{equation}
\label{Abar=0}
\bar{A}_\mu(x) = 0,~ \forall x \in  \text{Newtonian and strong regimes of gravity}\,.
\end{equation} 
Eq. \eqref{Abar=0} suggests  that $\bar{A}$ is a  composite field:
\begin{equation}\label{defineA}
\bar{A}_\mu\equiv \sum_{\alpha=1}^n B_\mu^\alpha\,,
\end{equation}
where $n$ is a natural number. We assume that  $B_\mu^{\alpha}$ are  gauge fields.  The observed dynamics of gravity in the strong and Newtonian regimes  then fixes the action of metric to the Einstein-Hilbert action:
\begin{equation}\label{Sg}
S[g] \equiv \frac{1}{16 \pi G} \int d^4 x \sqrt{- \det g} R\,,
\end{equation}
where $R$ is the Ricci scalar constructed out from metric  and its derivatives.   We set the dynamics of metric in all regimes of gravity to \eqref{Sg}. 

Now consider a set of particles $m_i, i\in \{1, N\}$. Assume that the manifold of space-time is smooth such that   the world-line of these particles can be parametrized with a global choice of time. Their actions follows from \eqref{s0}:
\begin{equation}\label{sdis}
S_M\equiv   \int^p_q dt \left( \sum_{i=1}^N   \frac{m_i}{2} g_{\mu\nu} \dot{x}^\mu_i \dot{x}^\nu_i\,-\sum_{i} m_i \bar{A}_\mu  \dot{x}_i^\mu - \sum_{i \neq j} V(x_i, x_j) \right)\,,
\end{equation}
where the last term describes possible non-gravitational interactions between the particles. In the continuum approximation then  \eqref{sdis} is mapped to 
\begin{equation}\label{SMatter}
S_M  =\int d^4 x \sqrt{-\det g } \left(\frac{1}{2} \rho g_{\mu\nu} u^\mu u^\nu - \bar{A}_\mu \rho u^\mu - \int d^4y \sqrt{-\det g} V(x,y)\right) \,,
\end{equation}
where $\rho$ is the density and $u^\mu$ is the four velocity vector field. Eq. \eqref{SMatter} is the matters' action of the GVT gravity. It also says how the gauge fields $B_\mu^\alpha$ are coupled to  $
J^\mu\equiv \rho u^\mu$.

\section{Gauge fields' action}
We  investigate the case wherein the gauge fields' action is a functional of the field strength of the gauge fields and their derivatives but  minimally coupled to  metric:
\begin{eqnarray}
S_{\text{GF}} &\equiv& S[\{B_{\mu\nu}^\alpha, \partial_\mu\}, g_{\mu\nu}]\,, \\
B_{\mu\nu}^\alpha &\equiv& \partial_\mu B_\nu^\alpha - \partial_\nu B_\mu^\alpha\,.
\end{eqnarray}
We consider  no mixing between the gauge fields. We demand that the equations of motion of the gauge fields to be second order partial derivative. These set 
\begin{eqnarray}
S_{\text{GF}} &=& \sum_{\alpha=1}^n S_\alpha[B_{\mu\nu}^\alpha, g_{\mu\nu}]\,.
\end{eqnarray}
We take into account the following simple choice for $S_\alpha$:
\begin{eqnarray}
\label{Sgauge}
S_\alpha &=&  -\frac{1}{16 \pi G k_\alpha l_\alpha^2} \int d^4x \sqrt{-\det g}~ {\cal L}_{\alpha}\left(\frac{l_\alpha^2}{4} B_{\mu\nu}^\alpha B^{\mu\nu}_\alpha\right),
\end{eqnarray}
where $k_\alpha$ is the coupling of the gauge field, $l_\alpha$ is the dimension-full parameter associated to the gauge fields and ${\cal L}_{\alpha}$ is the Lagrangian density. For sake of simplicity, we also set
\begin{equation}
\forall \alpha \,~~{\cal L}_\alpha (x) \equiv {\cal L} (x)\,  \,.
\end{equation}

\section{Regimes of the GVT gravity}
The set of \eqref{Sg},\eqref{SMatter} and \eqref{Sgauge} defines the action of GVT gravity. Taking the variation of this action with respect to $B_\alpha$ yields:
\begin{equation}\label{EB}
\nabla_\nu ({\cal L}'_\alpha B^{\nu\mu}_\alpha) = 16 \pi k_\alpha J^\mu\,,
\end{equation}
where ${\cal L}'_\alpha= \frac{d{\cal L}_\alpha(x)}{dx}$. 
Impose the following MONDian asymptotic behaviors on $\cal L$: 
\begin{equation}\label{Ltot}
{\cal L}'(x)\,=\,
\left\{
\begin{array} {ccllr}
1 &,&~ |x|\gg1&~~\text{Regime a}
\\
|x|^{1/2}&,&~|x|\le 1&~~\text{Regime b}
\end{array}
\right.\,.
\end{equation}
Sufficiently large values for the field strengths are in the regime a of \eqref{Ltot} for which \eqref{defineA} and  \eqref{EB} result
\begin{equation}
\nabla_\nu \bar{A}^{\nu\mu} = \sum_{\alpha=1}^n \nabla_\nu B^{\nu\mu}_\alpha = 16 \pi (\sum_{\alpha=1}^n k_\alpha) J^\mu\,,
\end{equation}
where $\bar{A}^{\nu\mu}$ stands for the field strength of $\bar{A}^\mu$.  The consistency with \eqref{Abar=0} then requires 
\begin{equation}\label{sk=0}
\sum_{\alpha=1}^n k_\alpha \equiv 0\,.
\end{equation} 
This means that we should refer to the regime a of \eqref{Ltot} the strong and Newtonian regimes of GVT theory. 

Eq. \eqref{EB} indicates that strong and Newtonian  regimes are generally around and sufficiently close to the mass distribution. The net contribution of the gauge fields to the energy momentum tensor is vanishing in the strong and Newtonian regimes. The strong and Newtonian regimes of GVT and Einstein-Hilbert action, therefore, are identical at the level of action and the equations of motion. This identically implies  that the strong and the Newtonian regimes of GVT are as stable as  those of the Einstein-Hilbert action.  

The identically furthermore implies that the metric in the Newtonian regime  of a localized slow-moving mass distribution in an asymptotically flat space-time is
\begin{equation}\label{metric}
g_{\mu\nu} dx^\mu dx^\nu = -(1+2 \phi_N) dt^2 + (1-2 \phi_N) \delta_{ij} dx^i dx^j\,,
\end{equation}
where $\phi_N$ is the Newtonian potential:
\begin{equation}\label{phiN}
\nabla^2 \phi_N = 4 \pi G \rho\,,
\end{equation}
where $\rho$ is the density.  The gauge fields in the Newtonian regime follow from  
\begin{equation}
J^\mu= \rho \delta^\mu_{0}\, \to B^\alpha_\mu =  \delta_\mu^0 \phi^\alpha\,\,\label{48}\,,
\end{equation}
and utilizing \eqref{Ltot} beside expressing the solution in term of the solution of \eqref{phiN}:
\begin{equation}\label{Bsol}
\phi^\alpha = 4  k_\alpha  \phi_N\,.
\end{equation}
Eq. \eqref{Bsol} says that as we move away from the center of the mass distribution the gauge field strength decreases. As the gauge field strength decreases we start moving toward the regime b of \eqref{Ltot}.  
Note that the gauge fields contribute to  the energy momentum tensor in the regime b, however,  the  regime b occurs sufficiently far away from the mass distribution. The contribution is, therefore, negligible and  \eqref{metric} describes the leading order metric also in the  regime b.   In order to find the gauge fields in the regime b insert  \eqref{48} into \eqref{EB}:
\begin{subequations}
\label{aqualeq}
\begin{eqnarray}
\nabla^i \left(\frac{|\nabla \phi^\alpha|}{a_\alpha} \nabla_i \phi^\alpha \right) &=& \text{sign}(k_\alpha) 4 \pi G  \rho\,,\\
a_\alpha& =&  \frac{4 \sqrt{2} |k_\alpha|}{l_\alpha} c^2\,,
\end{eqnarray}
\end{subequations}
where the explicit dependency on the light speed is recovered. The solution to \eqref{aqualeq}  can be expressed in term of the solution of the AQUAL's theory \cite{AQUAL} with the critical acceleration of $a_\alpha$, $\phi^A(x,a_\alpha)$:
\begin{equation}\label{Bbsol}
\phi^\alpha = \text{sign}(k_\alpha) \phi^A(x,a_\alpha)\,.
\end{equation}
Negative $k_\alpha$ may lead to dynamical instability in the MOND regime.  To prevent these instabilities we consider  the limit of   $ l_\alpha \to \infty$ for $k_\alpha < 0$.  So dynamical instabilities are prevented in the MOND regime. This concludes solving the equations of motion for the metric and gauge fields. 

\section{Orbits of particles}
Eq. \eqref{s0} describes the effective action for a massive test particle in GVT gravity.  In its strong and Newtonian regimes it coincides to that of the Einstein-Hilbert action due to  \eqref{sk=0}. In the Newtonain and regime b of \eqref{Ltot} around a slow-moving mass distribution, where \eqref{metric}  and \eqref{48} hold,   the effective action follows:
\begin{equation}\label{eft}
S_{GVT} = \int d\tau L = m \int d\tau \left(  -\frac{1+2 \phi_N}{2} \dot{t}^2 + \frac{1-2 \phi_N}{2} \delta_{ij} \dot{x}^i \dot{x}^j -\bar{\phi}  \dot{t} \right)\,,
\end{equation}
where 
\begin{equation}
\bar{\phi}\equiv \sum_{\alpha=1}^{n}\phi_\alpha\,,
\end{equation}
where  either  \eqref{Bsol} or \eqref{Bbsol} represents  $\phi_\alpha$. A very slow moving test particle 
\begin{subequations}
\begin{eqnarray}
\tau &=& t \,,\\
\frac{1-2 \phi_N}{2} \delta_{ij} \dot{x}^i \dot{x}^j &\approx& \frac{1}{2}\delta_{ij} \dot{x}^i \dot{x}^j \,,
\end{eqnarray} 
\end{subequations}
 simplifies \eqref{eft}: 
\begin{equation}\label{eff1}
S_{GVT} = \int d\tau L = \int dt \left(   \frac{m}{2} \delta_{ij} \dot{x}^i \dot{x}^j -m( \phi_N +\bar{\phi})  \right)\,,
\end{equation}
which is the action of a particle of mass $m$ in the effective gravitational field of $\phi_N +\bar{\phi}$.  It could have been derived from 
\begin{equation}\label{effDm1}
S_{DM} = \int d\tau L = m \int d\tau \left(  -\frac{1+2 (\phi_N+ \bar{\phi})}{2} \dot{t}^2 + \frac{1-2 (\phi_N+\bar{\phi}) }{2} \delta_{ij} \dot{x}^i \dot{x}^j  \right)\,,
\end{equation}
which represents the effective action of a massive test particle in the $\Lambda$CDM standard model provided thar ${\phi}_N$ and $\bar{\phi}$  are interpreted as the contribution of the baryonic and dark matter mass distribution to the geometry. 

In the Newtonian regime  (wherein \eqref{Bsol} represents  $\phi_\alpha$)  holds $\bar{\phi}=0$ due to \eqref{sk=0}. Therefore, eq. \eqref{eff1} coincides to the Newtonian dynamics. Now relabel the gauge fields such that  the regime b  first occurs for $\phi^1$.  Set:
\begin{subequations}
\label{B1}
 \begin{eqnarray}
 k_1&>&0\,,\\
 a_1 &=& a_0 = (1\pm 0.2) \times 10^{-10} \frac{m}{s^2}\,,
 \end{eqnarray} 
 \end{subequations}
 where $a_0$ is the critical acceleration of  MOND  \cite{MOND}.  \eqref{B1} enables $\phi_1$ to reproduce the MONDian dynamics in the regime b. We refer to the regime b as the MOND regime. One can set th  the parameters  of the rest of the gauge fields such that the MOND regime of $B^1_\mu$ and the Newtonian regimes of the rest of the gauge fields reproduce the observed MONDian behavior around spiral galaxies. This is possible in case we have at least two gauge fields.  This proves that GVT gravity is capable of  reproducing the flat rotational velocity curves and  the Tully-Fisher relation \cite{Tully}.  Let it be highlighted  the MONDian behavior would not propagate to infinity for
\begin{equation}\label{5.7}
\sum_{\alpha=1}^n \text{sign}(k_\alpha) \sqrt{a_\alpha}=0
\end{equation} 
due to  asymptotic behavior of $\phi^A(x,a_\alpha)$ at infinity. Since we have secured stability of the theory  in the line after eq. \eqref{Bbsol} then \eqref{5.7} can not be satisfied.  The MONDian regime of a stable GVT theory propagates to infinity. 
To describe  the orbit of a fast moving particle (such as photon),  express $t$ and $x_1$ in term of Dirac coordinates \cite{Dirac}  
\begin{eqnarray}
t &=& x_++x_- \,,\\
x_1 & = & x_+-x_-\,, 
\end{eqnarray}
and choose the affine parameter of $\tau=x_+$ and use the approximation of $\dot{x_2}, \dot{x_3}, \dot{x_-}\ll1$. These simplify \eqref{eft} to  
\begin{equation}\label{eff2}
S_{GVT} =  m \int dx_+ \left( \frac{1}{2} ( \dot{x}_2^2+\dot{x}_3^2) - 2 \dot{x}_-  - ( 2 \phi_N +\bar{\phi})  \right)\,,
\end{equation}
Applying the same procedure upon a metric with an arbitrary value for the post Newtonian parameter of $\gamma$ 
\begin{equation}\label{effDm22}
S_{\gamma} =  m \int d\tau \left(  -\frac{1+2 (\phi_N+ \bar{\phi})}{2} \dot{t}^2 + \frac{1-2 \gamma (\phi_N+\bar{\phi}) }{2} \delta_{ij} \dot{x}^i \dot{x}^j  \right)\,,
\end{equation}
 gives 
\begin{equation}\label{eff3}
S_{\gamma} =  m \int dx_+ \left( \frac{1}{2} ( \dot{x}_2^2+\dot{x}_3^2) - 2 \dot{x}_-  - (1+\gamma)( \phi_N +\bar{\phi})  \right)\,,
\end{equation}
Eq.  \eqref{eff2} and \eqref{eff3}  shows that the physical $\gamma$ parameter of the GVT theory  vanishes in the MOND regime of the theory: $\gamma_{GVT}=0$.  Let it be emphasized that there exists no data available on the post Newtonian parameter of  $\gamma$ in the MONDian regimes, in regimes wherein gravitational acceleration is smaller than $a_0$ \cite{PNP}.  $\Lambda$CDM assume that it holds $\gamma=1$ while GVT gravity requires $\gamma=0$ in the MONDian regimes. This suggests to observationally measure the $\gamma$ parameter in the MOND regimes in order to decide which theory better fits the data.

 \providecommand{\href}[2]{#2}\begingroup\raggedright
  \end{document}